\documentclass[english,keywords,showpacs,amsmath,amssymb,twocolumn]{revtex4}
\usepackage[T1]{fontenc}
\usepackage[latin1]{inputenc}
\usepackage{babel}
\usepackage{array}
\usepackage{graphics}
\usepackage{subfigure}
\makeatother
\begin{document}
\title{Contextuality within quantum mechanics  manifested in subensemble mean values}
\author{Alok Kumar Pan\footnote{apan@bosemain.boseinst.ac.in}$^1$,
and Dipankar Home\footnote{dhome@bosemain.boseinst.ac.in}$^1$}

\affiliation{$^1$ CAPSS, Department of Physics, Bose Institute, Sector-V, Salt Lake, Calcutta
700091, India}
\begin{abstract}
For spin-1/2 particles, using a suitable Mach-Zehnder-type setup with a spin-flipper, we argue that it is a direct consequence of the quantum mechanical treatment that an experimentally verifiable \textit{subensemble} mean of the measured values of an arbitrarily chosen spin variable exhibits dependence on the choice of a comeasurable `path' observable. This, in turn, enables inferring path-spin contextuality at the level of individual measured values of spin that  are predetermined using a relevant hidden-variable model applied to our setup.
\end{abstract}
\pacs{03.65.Ta}
\maketitle
\section{Introduction}
In quantum mechanics (QM), the expectation value of any dynamical variable is fixed by a given wave function, and is independent of the experimental setup used for measuring the variable in question. In particular, this value is independent of the measurement (previous or simultaneous) of any other commuting (comeasurable) observable. The assumed extension of such \textit{`context-independence'} from the level of quantum expectation values to any \textit{individual} measured value of a dynamical variable that is \textit{predetermined} is what underpins a class of theories known as the deterministic noncontextual hidden-variable (NCHV)models. Such realist models seek to provide event-by-event descriptions of  quantum phenomena by using a `more complete' specification of the state of an individual system than that is given by a wave function. The study of the issue concerning compatibility between the formalism of QM and the NCHV models has been one of the active areas of investigations related to the foundational issues of QM.

Different versions of the proof of the mathematical theorem showing an incompatibility between the formalism of QM and the deterministic NCHV models were given by Gleason \cite{gleason}, Bell \cite {bell}, Kochen and Specker \cite{kochen}, followed by others suggesting a variety of ingenious proofs of this `no-go theorem'; see, for example, \cite{mermin, peres, cabello, penrose}. But, much later, an important observation was made\cite{cabello2} that since all these proofs used some inputs from the formalism of QM, and were thus \textit{not} entirely independent of the formal structure of QM, they could not be used for \textit{experimentally discriminating} between QM and the NCHV models.

The above point, though, was taken into account in an earlier study by Roy and Singh \cite{roy}  while showing an incompatibility between QM and a Bell-type inequality obtained from the \textit{stochastic} NCHV models (in which for every possible value of the hidden variable, the \textit{probability} to find a certain result for a given observable does not depend on which other observables are measured jointly, and the joint probabilities are simply products of the probabilities for single observables) for particles with spin higher than $1/2$. For the \textit{deterministic} NCHV models, Cabello and Garcia-Alcaine \cite{cabello2} gave an argument using a two-particle two-state system that enables a suitable joint measurement pertaining to a particular set of compatible propositions to discriminate between QM and a testable consequence of the NCHV hypothesis that is derived independent of QM. Although this led to some interesting work \cite{simon}, a ticklish point is that this type of \textit{non-statistical} argument in terms of the yes-no validity of propositions is contingent upon the relevant dynamical variables being measured with infinite precision - an issue that has been the subject of considerable discussions  \cite{meyer} as to what extent the \textit{finite precision} (in the sense of `imprecision' in actually what is being measured) measurements can enable an empirical discrimination between QM and the deterministic NCHV models.

In the context of the aforesaid controversy, a testable example was formulated by using an entanglement between the path and spin observables of a spin-$1/2$ particle so that a testable Bell-type inequality can be derived valid for any deterministic NCHV model,  but which is independent of QM \cite{home}. The feature that, in such a scheme, QM violates the noncontextual realist Bell-type inequality by a \textit{finite amount} suggests the possibility of empirically discriminating the NCHV models from QM, even if the actual measurements are inevitably imprecise.

The experimental realization of a variant of the above scheme was achieved by Hasegawa \emph{et. al.}\cite{hasegawa1} using single-neutron interferometry, and the NCHV models were refuted through an observed violation of the relevant Bell-type inequality. Thereafter, the experimental investigation along this line was enriched by more studies \cite{hasegawa2}. A significant recent development is the formulation of a scheme \cite{cabello4} for testing the NCHV models on the basis of the quantum mechanical violation of an inequality obtained from the Peres-Mermin\cite{mermin,peres} proof of the `no-go theorem'.This proposed experiment using neutrons has the potentiality of demonstrating the quantitatively largest violation of a bound expected from any NCHV model.

The whole body of the preceding works may, thus, be viewed as an enterprise that seeks to establish, with an increasing rigour, the untenability of the NCHV models in a way distinct from the studies related to local hidden-variable models\cite{zela}. Thus, these works contribute towards countering the doubts that have been raised (reviewed, for instance, by Barrett and Kent \cite{barrett}) about a decisive empirical refutation of the NCHV models that can be achieved by the \textit{finite precision} measurements. 

Against the above backdrop, the present Letter explores the issue of quantum contextuality from an entirely different perspective. Note that all the relevant studies have so far   focused on showing an incompatibility between QM and the NCHV models by deriving consequences from the basic tenets of the NCHV models and by showing their incompatibility with the corresponding implications of the formalism of QM. In contrast, in this Letter, we argue that it is indeed possible to derive, \textit{within} the formalism of QM, a statistically discernible effect of `contextuality' that is manifested in terms of the measured values  of a spin variable pertaining to either of the two \emph{subensembles}  comprising the final output ensemble. The specific sense in which the term `contextuality' is considered here pertains to the feature that the operationally well-defined subensemble spin mean values are contingent upon what \emph{choice} is made of measuring a suitably  defined comeasurable(commuting)`path' observable. We may stress that this form of contextuality has so far remained entirely unexplored.

Our demonstration is with respect to a specific setup that makes a suitable use of the \emph{path-spin entanglement} (that can be  labelled as `intraparticle entanglement') between the spin variables and the `path' observables of a spin-$1/2$ particle. Importantly, this effect is obtained by \textit{preserving} the context-independence of the quantum expectation value of spin that is defined with respect to the \textit{whole ensemble} of particles on which the measurements are performed. 

Further, a key aspect is that the quantum mechanically calculated `context-dependence' at the level of subensemble statistics enables the inference of `context-dependence' of an individual measured value of a spin variable that is predetermined using any hidden-variable model relevant to our setup. Thus, what is revealed is that the property of contextuality can be viewed as the one embodied \textit{within} the formalism of QM.

Now, before formulating our argument, we will first explain the specifics of the setup required. This discussion, though presented in terms of the spin-1/2 particles(such as neutrons), works equally well for photons with appropriate polarizing and analyzing devices. Interestingly, although such a setup was discussed earlier\cite{home, hasegawa1}, the implication brought out in this Letter has remained hitherto unnoticed.

\section{The setup}
Let us consider that an ensemble of neutrons, all corresponding to an initial spin polarized state along the $+\widehat {z}-axis$(denoted by $\left|\uparrow\right\rangle_{z}$) be incident on a  50:50 beam-splitter(BS1). An incident particle can then emerge along either the transmitted or the reflected channel corresponding to $\left|\psi_{1}\right\rangle$ or $\left|\psi_{2}\right\rangle$ respectively. Subsequently, the two beams corresponding to $\left|\psi_{1}\right\rangle$ and $\left|\psi_{2}\right\rangle$ are recombined at a second beam-splitter(BS2) whose reflection and transmission probabilities are $|\gamma|^{2}$ and $|\delta|^{2}$ respectively. 

Now, note that for any given lossless beam-splitter, arguments using the unitarity condition show that the phase shift between the transmitted and the reflected states of particles is essentially $\pi/2$\cite{zeilinger}. Further, in our treatment, we will henceforth take the reflection and transmission amplitudes of BS2 to be real quantities. 

%\vskip -2.1cm
In order to formulate our argument, it is necessary to introduce the mutually orthogonal `path' states $\left|\psi_{1}\right\rangle$ and $\left|\psi_{2}\right\rangle$ which are  eigenstates of the projections operators $P(\psi_{1})$ and $P(\psi_{2})$ respectively. These projection operators  correspond to the observables that pertain to the determination of \textit{`which channel'} a particle is found to be in. For example, the results of such  a measurement for the transmitted(reflected) channel with binary alternatives are given by the eigenvalues of $P(\psi_{1})$($P(\psi_{2})$); the eigenvalue $+1(0)$ corresponds to a neutron being found(not found)in the channel represented by $\left|\psi_{1}\right\rangle$($\left|\psi_{2}\right\rangle$). 

Next, we consider that the neutrons which move along one of the channels, say, the one corresponding to $|\psi_{1}\rangle$ pass through a spin-flipper(SF) (that contains a uniform magnetic field along the $+\widehat x$-axis) that flips the state $\left|\uparrow\right\rangle_{z}$ to $\left|\downarrow\right\rangle_{z}$. Subsequently, the neutrons passing through the channels $\left|\psi_{1}\right\rangle$ and $\left|\psi_{2}\right\rangle$ are reflected by the mirrors M2 and M1 respectively - these reflections do not lead to any net relative phase shift between $\left|\psi_{1}\right\rangle$ and $\left|\psi_{2}\right\rangle$. 

\begin{figure}[h]
{\rotatebox{0}{\resizebox{9.0cm}{7.0cm}{\includegraphics{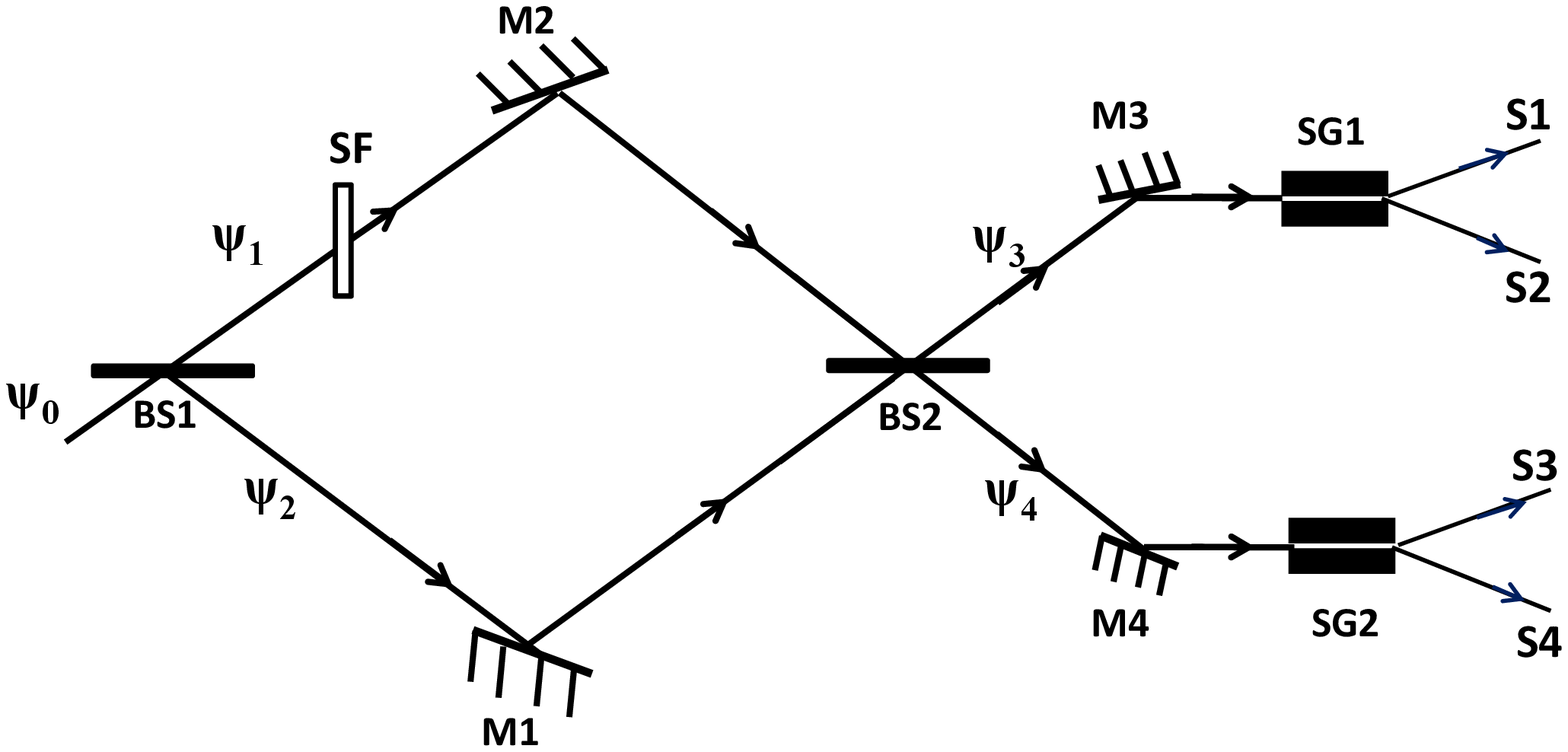}}}}
\vskip -2.2cm
\caption{\footnotesize Spin-1/2 particles (say, neutrons) entering this Mach-Zehnder type setup through a beam splitter BS1 pass through the channels corresponding to  $|\psi_{1}\rangle$ and $|\psi_{2}\rangle$. A spin-flipper(SF) is placed along one of the channels $|\psi_{1}\rangle$. Subsequently, spin measurements are considered on the neutrons emerging from the beam splitter BS2 by using the two spatially separated Stern-Gerlach devices SG1 and SG2. For inferring quantum contextuality, \emph{subensemble mean values} of the spin variable $\widehat\sigma_{\theta}$ are considered pertaining to SG1 and SG2 separately.}
\end{figure}

Thus, for the neutrons with spin $\left|\uparrow\right\rangle_{z}$ incident on BS1, the state of those incident on BS2 is represented by
\begin{eqnarray}
\left|\Psi\right\rangle=\frac{1}{\sqrt{2}}\left(\left|\psi_{1}\right\rangle\left|\downarrow\right\rangle_{z}
+i \left|\psi_{2}\right\rangle\left|\uparrow\right\rangle_{z}\right)
\end{eqnarray}
where, in writing Eq.(1), we have taken into account a relative phase shift of $\pi/2$ between $\left|\psi_{1}\right\rangle$ and $\left|\psi_{2}\right\rangle$ that  arises because of the reflection from BS1.

Here we stress that the process of passing neutrons through the arrangement of BS1+SF serves the purpose of appropriately \emph{preparing} a state given by $|\Psi\rangle$ of Eq.(1) on which we will consider measurements relevant to our demonstration of contextuality. Note that the state thus prepared is an \emph{entangled state} involving the path degrees of freedom and the spin variables of a spin-$1/2$ particle. 

\section{The required measurements}
For analysing the relevant measurements on the state $|\Psi\rangle$ which is prepared by the setup \emph{preceding}  BS2, we write the state after emerging from BS2 as given by
\begin{eqnarray}
\left|\Phi\right\rangle= \frac{1}{\sqrt{2}}\left[i\left|\psi_{3}\right\rangle(\gamma \left|\downarrow\right\rangle_{z} + \delta   \left|\uparrow\right\rangle_{z})
+  \left|\psi_{4}\right\rangle\left(\delta\left|\downarrow\right\rangle_{z}-\gamma \left|\uparrow\right\rangle_{z}\right)\right]
\end{eqnarray}where the output states $\left|\psi_{3}\right\rangle$ and $\left|\psi_{4}\right\rangle$ are unitarily related to the  states $\left|\psi_{1}\right\rangle$ and $\left|\psi_{2}\right\rangle$ by the following relations
\begin{eqnarray}
\left|\psi_{1}\right\rangle = i\gamma \left|\psi_{3}\right\rangle+ \delta  \left|\psi_{4}\right\rangle\\
\nonumber
\left|\psi_{2}\right\rangle = i\gamma \left|\psi_{4}\right\rangle+ \delta \left|\psi_{3}\right\rangle
\end{eqnarray}
where $\gamma$ and $\delta$ satisfy  $\gamma^{2}+\delta^{2}=1$.

Eqs.(2) and (3) show that, for a given linear combination of  $\left|\psi_{1}\right\rangle$ and $\left|\psi_{2}\right\rangle$, using the different values of  $\gamma$($\delta$), one can generate at the output of BS2, various linear combinations of $\left|\psi_{1}\right\rangle$ and $\left|\psi_{2}\right\rangle$ that correspond to different probability amplitudes of finding particles in the channels corresponding to $\left|\psi_{3}\right\rangle$ and $\left|\psi_{4}\right\rangle$. Then, the different values of the BS2 beam-splitter parameter $\gamma(\delta)$ can be regarded as corrresponding to different \emph{choices} of what may be called the `path' observables $\widehat{A}_{i}$. Thus, BS2 plays a key role as a part of this measuring arrangement.

Formally, one can write $\widehat{A}_{i}=P(\psi_{3})-P(\psi_{4})$ where the eigenvalues $\pm 1$ of $\widehat{A}_{i}$ pertain to the detection of a particle in a channel corresponding to either $\left|\psi_{3}\right\rangle$ or  $\left|\psi_{4}\right\rangle$ respectively. An important point is that there is an \emph{isomorphism}\cite{takasaki}  between the algebra of the observables $\widehat{A}_{i}$ and the algebra of $2\times 2$ complex matrices spanned as a linear space by the Pauli matrices $\sigma_{x}$, $\sigma_{y}$, $\sigma_{z}$, and the identity $I$ matrix. This feature can be described as follows.

 Taking the representation, for instance, 
$|\psi_{1}\rangle  \rightarrow \left(\begin{array}{c} 
    1\\    0   \end{array}\right)$ , 
    \hskip 0.3cm 
    $|\psi_{2}\rangle \rightarrow \left(\begin{array}{c}   0\\  1    
     \end{array}\right)$     
     and, using the relations given by Eq.(3), it follows that
\begin{equation}
A_{i}= \left(\begin{array}{cl} 
    \gamma^{2}-\delta^{2}& \ -i 2\gamma\delta \\  i 2\gamma\delta  & \ \delta^{2}- \gamma^{2} \end{array}\right) 
\end{equation} 
that can be written as the following linear combination of the Pauli matrices 
\begin{equation}
A_{i}=2\gamma\delta \sigma_{y}+ (\gamma^{2}-\delta^{2}) \sigma_{z}= \vec{\sigma}.\vec{a}_{i}
\end{equation}
where $\vec{a}_{i}= 2\gamma\delta \widehat{j}+ (\gamma^{2}-\delta^{2}) \widehat{k}$

Eq.(5) shows that, given a particular choice of the beam-splitter parameter $\gamma(\delta)$, a specific  `path' observable $A_{i}$  can be viewed as corresponding to a definite vector component defined by $\vec{\sigma}.\vec{a}$. It is in this sense that such `path' observables  may be regarded as `\emph{pseudo}-spin' observables that commute with the actual spin observables.

\section{The argument inferring quantum contextuality} 
We focus on the measurement of an arbitrary spin variable, say, $\widehat\sigma_{\theta}$ with the relevant outcomes being registered by the two suitably oriented Stern-Gerlach(SG1 and SG2) devices(Fig.1)placed along the spatially separated channels $|\psi_{3}\rangle$ and $|\psi_{4}\rangle$ respectively, while the counts registered at SG1 and SG2 also yield information about the outcomes pertaining to the measurement of the `path' observable $\widehat A_{i}$. Note that the eigenstates of $\widehat\sigma_{\theta}$ can be written as $\left|\uparrow\right\rangle_{\theta}= cos\theta\left|\uparrow\right\rangle_{z}+ sin\theta\left|\downarrow\right\rangle_{z}$ and $\left|\downarrow\right\rangle_{\theta}= sin\theta\left|\uparrow\right\rangle_{z} - cos\theta\left|\downarrow\right\rangle_{z}$.  Then, corresponding to the prepared path-spin entangled state $\left|\Psi\right\rangle$ given by Eq.(1), the expectation value of the spin variable $\widehat\sigma_{\theta}$ as measured by considering the \emph{whole ensemble} of particles emerging from the beam-splitter BS2(part of the measuring setup here) is of the form

\begin{equation}
\langle\widehat{\sigma}_{\theta}\rangle_{\Psi}=0
\end{equation}

Here we would like to stress that the measurement of the above expectation value involves contributions from both the output subensembles  corresponding to the counts \emph{separately} registered in the measuring devices SG1 and SG2. The respective \emph{subensemble spin mean values}, calculated from $|\Phi\rangle$ given by Eq.(2), are denoted by $(\bar{\sigma}_{\theta})_{SG1}$ and $(\bar{\sigma}_{\theta})_{SG2}$, whence 
\begin{equation}
\left\langle \widehat{\sigma}_{\theta}\right\rangle_{\Psi}=(\bar{\sigma}_{\theta})_{SG1}+ (\bar{\sigma}_{\theta})_{SG2}
\end{equation} 
Note that all the three quantities occurring in the equality given by Eq.(7) have the same operational status as far as their \emph{statistical reproducibility} is concerned. But there is a crucial distinction between the left and the right hand sides of Eq.(7) with respect to the issue of path-spin interdependence. The quantity on the left  hand side of Eq.(7), the spin expectation value $\langle\widehat{\sigma}_{\theta}\rangle_\Psi$ pertaining to the whole ensemble, is \emph{independent} of \textit{which} `path' observable is measured along with it. On the other hand, if we consider  the quantities on the right hand side of Eq.(7), each of the subensemble spin mean values $(\bar{\sigma}_{\theta})_{SG1} $ and $(\bar{\sigma}_{\theta})_{SG2} $  is found to be sensitive to the \emph{choice} of the comeasurable `path' observable.  This can be verified by considering the spin measurement outcomes pertaining to each of the output subensembles. 

For this, we proceed by noting that, using Eq.(2), we have for the respective subensemble spin mean values

\begin{eqnarray}
(\bar{\sigma}_{\theta})_{SG1}=\frac{1}{2}(\delta^{2}-\gamma^{2}) cos 2\theta + \gamma\delta sin2\theta\\
(\bar{\sigma}_{\theta})_{SG2}=\frac{1}{2}(\gamma^{2}-\delta^{2})cos2\theta - \gamma\delta sin2\theta
\end{eqnarray}

Next, we come to the crux of our argument that hinges on two different choices of $\gamma$($\delta$) and where the superscript $A_{1}$($A_{2}$) is used to specify the choice of the `path' observable that can be measured in a given context: 

$(a)$ Taking $\gamma=\delta=1/\sqrt{2}$,  this choice implies  the measurement of a particular  `path' observable $A_{1}=\vec{\sigma}.\vec{a}_{1}$ where $\vec{a}_{1}=\widehat{j}$. In this case, using Eqs.(8-9), we obtain for SG1 and SG2 separately 

\begin{eqnarray}
(\bar{\sigma}_{\theta})^{(A_1)}_{SG1}=\gamma\delta sin2\theta; \hskip 0.3cm 
(\bar{\sigma}_{\theta})^{(A_1)}_{SG2}= -\gamma\delta sin2\theta
\end{eqnarray}

whence, using Eq.(2), the spin expectation value for the whole ensemble is given by $\langle\widehat{\sigma}_{\theta}\rangle_{\Psi}=0$.

$(b)$ Taking $\gamma=1/2$ and $\delta=\sqrt{3}/2$, this choice  implies the measurement of a \emph{different} `path' observable $A_{2}=\vec{\sigma}.\vec{a}_{2}$ where $\vec{a}_{2}= \frac{\sqrt{3}}{2} \widehat{j}- \frac{1}{2} \widehat{k}$.  Consequently, using Eqs.(8-9),we obtain

\begin{eqnarray}
\nonumber
(\bar{\sigma}_{\theta})^{(A_2)}_{SG1}= \frac{1}{4}cos2\theta +\frac{\sqrt{3}}{4}sin2\theta\\ 
(\bar{\sigma}_{\theta})^{(A_2)}_{SG2}= -\frac{1}{4} cos2\theta -\frac{\sqrt{3}}{4}sin2\theta
\end{eqnarray}
whence, using Eq.(2), the spin expectation value for the whole ensemble is given by  $\langle\widehat{\sigma}_{\theta}\rangle_{\Psi}=0$. 

It is then evident from Eqs.(10-11) that, while the spin expectation value $\langle\widehat{\sigma}_{\theta}\rangle_{\Psi}$ pertaining to the \emph{whole ensemble} remains the \emph{same} for both the choices of $A_{1}$ and $A_{2}$, the path-spin interdependence gets manifested in terms of the \emph{subensemble spin mean values} given by the  testable quantities   $(\bar{\sigma}_{\theta})^{(A_{1}, A_{2})}_{SG1}$ and $(\bar{\sigma}_{\theta})^{(A_{1}, A_{2})}_{SG2}$. To put it  precisely, in our example, the interdependence between the `path' and the spin variables has the following operational meaning

\begin{eqnarray}
(\bar{\sigma}_{\theta})^{(A_1)}_{SG1}\neq(\bar{\sigma}_{\theta})^{(A_2)}_{SG1}; \hskip 0.3cm 
(\bar{\sigma}_{\theta})^{(A_1)}_{SG2}\neq(\bar{\sigma}_{\theta})^{(A_2)}_{SG2}
\end{eqnarray}
 i.e., the subensemble mean value of the spin variable $\widehat{\sigma}_{\theta}$ depends upon \emph{which} of the `path' observables $\widehat{A}_{1}$ or $\widehat{A}_{2}$ is comeasured, where both $\widehat{ A}_{1}$ and $\widehat{A}_{2}$ commute with $\widehat{\sigma}_{\theta}$. 
 
A crucial point to be stressed is that the above manifestation of contextuality holds good \emph{whatever} be the choice of measuring the spin variable $\widehat{\sigma}_{\theta}$; i.e., this effect occurs for \emph{any} value of $\theta$, thereby underscoring the essentially quantum mechanical nature of the effect demonstrated here. This is because, while the path-spin entangled state is prepared in our setup such that it involves the spin-polarized state along a particular direction, viz. the $\widehat{z}$-axis, it is the quantum superposition of spin states with respect to any basis, used explicitly in writing Eqs.(8) and (9), that ensures the statistical display of path-spin interdependence for the measured values of any \emph{arbitrary} spin component $\widehat{\sigma}_{\theta}$ pertaining to each of the two subensembles of particles in the output channels. This effect is, thus, {\it not} reproducible using any classical model using pre-existing noncontextual values of dynamical variables. 

Further, note that although the above treatment has been given in terms of the subensemble averages of the measured spin values, it can be verified that the dependence on the choice of the comeasurable `path' observable manifests as well in terms of the relevant fluctuations $(\Delta\widehat{\sigma}_{\theta})=\sqrt{\langle \widehat{\sigma}^{2}_{\theta}\rangle - \langle \widehat{\sigma}_{\theta}}\rangle^{2}$ pertaining to the measured values of spins for the subensembles under consideration.

The above signature of contextuality is, thus, entirely a consequence of the quantum mechanical treatment of the setup considered here, obtained by  confining attention to the statistics of the spin measurement outcomes for \textit{each} of the two  subensembles at the exit channels. Note that, although the above argument is given specifically for the state preparation involving  a 50:50 beam-splitter BS1,  the treatment can also be shown to be valid for a general beam-splitter BS1, i.e., for different choices of reflectivity(transmittivity). Hence this demonstration is \emph{not} state-specific. 

Next, we remark that since at the exit channels, each of the subensemble spin mean values exhibits dependence on the choice of a comeasurable `path' observable $\widehat{A}_{1} (\widehat{A}_{2})$, this feature suggests that an \emph{individual} measured value of a spin variable which is predetermined using a relevant hidden-variable model applied to our setup would also depend on the comeasured `path' observable  $\widehat{A}_{1}$($\widehat{A}_{2}$) whose choice characterizes the context of this experiment. The meaning of such context-dependence within the framework of  a hidden-variable model that is applicable to our setup can be explained as follows.

Let us consider two distinct hidden-variables that determine separately the individual outcomes of the spin measurement in the output channels, and whether an individual particle is reflected/transmitted at the beam-splitter BS2 that occurs subsequent to the preparation of the path-spin entangled state on which the relevant measurements are considered. For a given prepared state, such hidden-variables be denoted by $\lambda$ and $\mu$ respectively where, for the sake of illustration, we stipulate that for, say, $\lambda=+1(-1)$, the particle is found in the up(down) channel of either of the SG devices that are used for measuring a spin variable, while for a given value of $\gamma(\delta)$ characterising the  beam-splitter BS2 , if, say, $\mu< \gamma^{2}(\mu\geq \gamma^{2})$, the particle is transmitted(reflected) at BS2. 

Now, if the parameter $\gamma(\delta)$ is varied, some of the particles earlier transmitted(reflected) will be reflected(transmitted), and, consequently, the distribution of the values of $\mu$ for the particles emerging in the reflected/ transmitted channel will vary. Concomitantly, the distribution of the values of $\lambda$ for reflected/transmitted particles must also change, thereby affecting the individual measured values of spin. Note that the latter feature is necessary for the compatibility with the quantum mechanical result that, as $\gamma(\delta)$ is varied, the statistical properties of the spin measurement results pertaining to each of the \textit{subensembles} registered at the respective SG devices SG1 and SG2 must also get changed. At the same time, however, this change of the distribution of $\lambda$ values for the reflected/transmitted particles, induced by the variation of $\gamma (\delta)$, needs to be such that the statistical results of spin measurements for the \emph{whole ensemble} of particles emerging from BS2 remain unchanged. 

Thus, the upshot of the above argument is that, in  terms of a relevant hidden-variable model, if one analyses the quantum mechanical effect of path-spin contextuality shown in this paper, a nontrivial \textit{correlation} would be required between the way the distributions of the values of the pertinent hidden-variables $\lambda$ and $\mu$ undergo changes as the BS2 parameter $\gamma(\delta)$ is varied. It is such correlation at the level of hidden-variables that embodies path-spin contextuality in the sense we have discussed for our example.The detailed features of such correlation  merit close scrutiny and should be an interesting direction for further study.

\section{Concluding remarks}
In sum, the argument presented here reveals a curious characteristic of the contextuality relation between the path and the spin degrees of freedom of a spin-1/2 particle that is \emph{distinct} from the contextuality involving other mutually commuting dynamical variables. While the former, as shown in this paper, is amenable to a quantum mechanically derivable  manifestation in terms of subensemble statistics, the latter is definable essentially in terms  of an incompatibility between any NCHV model and the QM formalism. Investigations are, thus, called for to gain insights into the nuances of this contrast between the path-spin contextuality and  the form of contextuality that exists between other mutually commuting dynamical variables; in particular, the difference between the type of contextuality shown here and the Kochen-Specker-Bell-type contextuality needs to be pinpointed. As to whether the effect demonstrated in the present paper also holds for other possible schemes that may implement joint path-spin measurement is an interesting question. Finally, a deeper understanding is required of the way in which, in the example considered in this Letter, the noncontextual character of QM is preserved for the  expectation value of spin defined for the whole ensemble, while the quantum formalism allows for the display of path-spin contextuality  in terms of the subensemble statistical results for the measured spin values.

\section*{Acknowledgments}
DH thanks H. Rauch and Y. Hasegawa for the useful interactions that served as a prelude to this work.  AKP recalls the discussions, in particular, with A. Cabello during his visit to Benasque Center for Science, Spain. AKP acknowledges the Research Associateship of Bose Institute, Kolkata.  DH thanks the DST, Govt. of India and Centre for Science and Consciousness, Kolkata for support. 

\end{document}